\documentstyle[12pt,epsf,aasms4]{article}

\begin{document}
\title{MACS: A quest for the most massive galaxy clusters in the universe}
\author{H.\ Ebeling\altaffilmark{1}, A.C.\ Edge\altaffilmark{2}, J.P.\ Henry\altaffilmark{1}}
\altaffiltext{1}{Institute for Astronomy, 2680 Woodlawn Drive, Honolulu,
       Hawaii 96822, USA}
\altaffiltext{2}{Department of Physics, University of
       Durham, South Road, Durham DH1\,3LE, UK}

\slugcomment{to appear in the June 1, 2001 issue of ApJ}

\begin{abstract}

We describe the design and current status of a new X-ray cluster
survey aimed at the compilation of a statistically complete sample of
very X-ray luminous (and thus, by inference, massive), distant
clusters of galaxies. The primary goal of the MAssive Cluster Survey
(MACS) is to increase the number of known massive clusters at $z>0.3$
from a handful to hundreds. Upon completion of the survey, the MACS
cluster sample will greatly improve our ability to study {\em
quantitatively} the physical and cosmological parameters driving
cluster evolution at redshifts and luminosities poorly sampled by all
existing surveys.

To achieve these goals we apply an X-ray flux and X-ray hardness-ratio
cut to select distant cluster candidates from the ROSAT Bright Source
Catalogue. Starting from a list of more than 5,000 X-ray sources
within the survey area of 22,735 square degrees we use positional
cross-correlations with public catalogues of Galactic and
extragalactic objects, reference to Automated Plate Measuring Machine
(APM) colours, visual inspection of Digitized Sky Survey images,
extensive CCD imaging, and finally spectroscopic observations with the
University of Hawaii's 2.2m and the Keck 10m telescopes to compile the
final cluster sample.

We discuss in detail the X-ray selection procedure and the resulting
selection function, and present model predictions for the number of
distant clusters expected to emerge from MACS.  At the time of this
writing the MACS cluster sample comprises 101 spectroscopically
confirmed clusters at $0.3\le z\le 0.6$; more than two thirds of these
are new discoveries. Our preliminary sample is already 15 times larger
than that of the EMSS in the same redshift and X-ray luminosity range.

\end{abstract}


\section{Introduction} 

The evolution of clusters of galaxies over cosmological timescales is
primarily driven by gravitational processes, such as the initial
gravitational collapse of overdense regions in the primordial universe
and their subsequent growth through accretion and cluster mergers. The
formation rate of the final products of this process -- relaxed,
massive galaxy clusters -- can be modeled straightforwardly for
different world models (Press \& Schechter 1976). The abundance of
clusters as a function of redshift is thus an important diagnostic of
cosmological parameters, primarily the normalized present-day matter
density of the universe, $\Omega_0$, and the amplitude of fluctuations
in that matter, $\sigma_8$ (e.g.\ Oukbir \& Blanchard 1997; Eke et
al.\ 1998; Henry 2000).

Although cosmological studies can, in principle, be conducted with
poor clusters, their slow evolution in all models of cluster formation
means that very large, statistically well defined samples at very high
redshift ($z\ga 1$) are required to obtain significant constraints. In
contrast, observations of the most massive systems, which are rarest
and evolve fastest in all cosmologies, provide tight constraints
already at moderate redshift. For instance, the predicted space
density of galaxy clusters with intra-cluster gas temperatures of
$kT\sim 7$ keV at $z\sim 0.5$ is more than a factor of ten higher in a
flat or open universe with $\Omega_0=0.3$ than in a closed universe
with $\Omega_0=1$; at $z\sim 1$ the difference approaches two orders
of magnitude (Viana \& Liddle 1996; Eke et al.\ 1996). For yet hotter
(i.e., more massive) clusters the dependance of the formation rate on
the chosen world model is even stronger.

As long as all systems are assumed to be virialized, only global
cluster properties (total X-ray luminosity, global gas temperature,
total mass) need to be known to constrain cosmological parameters.
Virialization is, however, only one and often an intermittent state,
preceded and, likely, interrupted by periods of growth through
mergers, accretion, and internal relaxation. A statistically complete,
large sample of massive, distant clusters would be invaluable to
investigate in detail the physical mechanisms governing these
evolutionary processes for the three main cluster components dark
matter, gas, and galaxies. Such an investigation is, again, most
feasible for massive clusters which are -- scatter in the respective
relations notwithstanding -- likely to be also the most X-ray luminous
and optically richest. They are therefore prime targets for studies of
the density and temperature distribution of the intra-cluster gas as
well as of the properties of the cluster galaxy population.  Galaxy
clusters also act as powerful gravitational lenses distorting the
images of background galaxies behind the cluster, and create
observable changes in the shape of the spectrum of the cosmic
microwave background (CMB) radiation passing through them
(Sunyaev-Zel`dovich [SZ] effect).  Lensing observations and detections
of the SZ effect allow independent measurements of the distribution of
dark matter and gas in clusters, and yet again the observed signal is
strongest for massive clusters.

In the local universe ($z\la 0.3$) dozens of massive clusters have
been known and studied in some detail for a long time. What we are
still lacking, and what is crucial for evolutionary studies, is a
sizeable sample of the high-redshift counterparts of these
well-studied local systems. In this paper we argue that the required
sample of massive, distant clusters is currently best compiled at
X-ray wavelengths, we present an overview of previous X-ray cluster
surveys, and show that the ROSAT All-Sky Survey can be used
efficiently to compile this sample (Section~\ref{xsurveys}). In
Section~\ref{macs} we introduce the MAssive Cluster Survey (MACS),
describe its characteristics and selection function, and discuss
predictions for the MACS sample size based on a no-evolution model.
Finally, we present a status report which demonstrates the efficiency
of our approach (Section~\ref{status}).

We assume $h=H_0/50$ Mpc s km$^{-1} = q_0 = 0.5$ throughout. Unless
explicitly stated otherwise, all X-ray fluxes and luminosities are
quoted in the 0.1--2.4 keV band.
 
\section{X-ray Cluster Surveys}
\label{xsurveys}

The arguably least biased and most secure way of detecting massive,
distant clusters is through wide-angle radio and sub-mm surveys
optimised to detect the Sunyaev-Zel`dovich (SZ) effect which is
independent of cluster redshift. However, with suitable SZ surveys
remaining infeasible for some time to come, the currently best way to
compile statistically complete cluster samples is through the
detection of X-ray emission from the hot intra-cluster gas. X-ray
cluster surveys are unbiased in the sense that they exclusively select
gravitationally bound objects and are essentially unaffected by
projection effects (e.g., van Haarlem, Frenk \& White 1997).  If
complete above a certain limiting X-ray flux, the resulting
statistical cluster samples will have a well-defined selection
function (a simple function of X-ray flux and, sometimes, X-ray
extent) that immediately allows the computation of the effective
survey volume for any real or hypothetical cluster.  Finally, an X-ray
cluster survey targeting only intrinsically X-ray luminous clusters
has the additional advantage of focusing on systems that are the ones
easiest to detect at any given redshift and for which the impact of
contamination from unresolved X-ray point sources is lowest.

\begin{figure}
\epsfxsize=\textwidth 
\hspace*{0mm} \epsffile{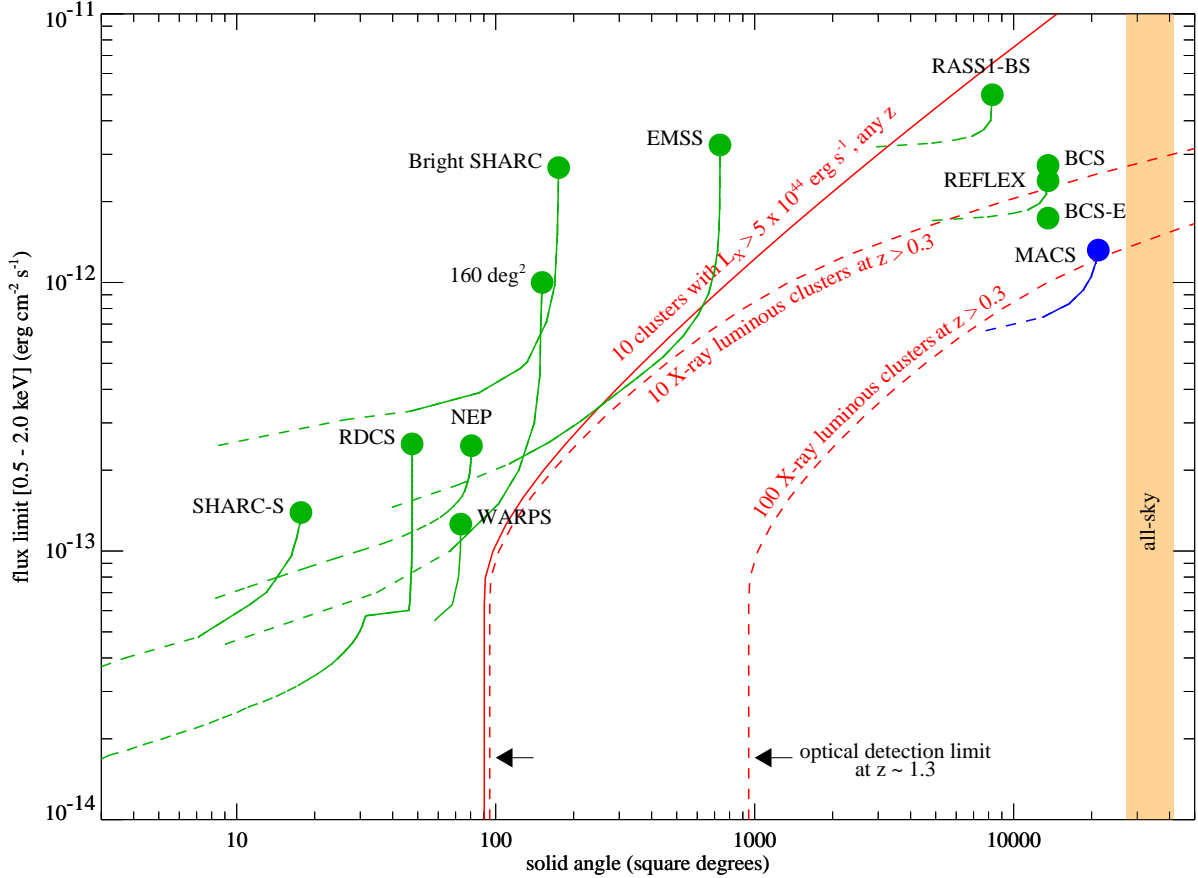} 
\caption{\small \label{surveys} The selection functions of all major
X-ray cluster surveys of the past decade. Also shown is the solid
angle required at a given flux limit to (statistically) detect 10 (or
100) X-ray luminous cluster at any redshift (or at $z>0.3$). Note how,
of all previous surveys, only the EMSS, BCS, and REFLEX projects are
just sensitive enough to detect a small number of distant, X-ray
luminous systems.}
\end{figure}

Several X-ray flux limited cluster samples have been compiled (and to
different degrees published) in the past decade; an overview of the
solid angles and flux limits of these surveys is presented in
Fig.~\ref{surveys}. Two kinds of surveys can be distinguished:
serendipitous cluster surveys (Bright SHARC, Romer et al.\ 2000a; CfA
160 deg$^2$ survey, Vikhlinin et al.\ 1998a; EMSS, Gioia et al.\
1990a; RDCS, Rosati et al.\ 1998; SHARC-S, Burke et al.\ 1997; WARPS,
Jones et al.\ 1998) and contiguous area surveys (BCS, Ebeling et al.\
1998; BCS-E, Ebeling et al.\ 2000a; NEP, Henry et al.\ 2001; RASS-BS,
DeGrandi et al.\ 1999; REFLEX, Guzzo et al.\ 1999). The former surveys
use data from pointed X-ray observations, whereas the latter are all
based on the ROSAT All-Sky Survey (RASS, Tr\"umper et al.\ 1993). With
the exception of the NEP survey, all contiguous cluster surveys cover
close to, or more than, 10,000 square degrees but are limited to the
X-ray brightest clusters. This fundamental difference in depth and sky
coverage has important consequences. As shown in Fig.~\ref{surveys},
the NEP survey as well as all serendipitous cluster surveys (with the
possible exception of the EMSS) cover too small a solid angle to
detect a significant number of X-ray luminous clusters (defined as
clusters with $L_{\rm X} > 5\times 10^{44}$ erg s$^{-1}$ in the
0.5--2.0 keV band or, equivalently, $L_{\rm X} > 8\times 10^{44}$ erg
s$^{-1}$ in the 0.1--2.4 keV band).  All previous RASS large-area
surveys, on the other hand, are capable of finding these rarest
systems, but are too shallow to detect them in large numbers at
$z>0.3$.

The observational situation summarized in Fig.~\ref{surveys} has led
to the misconception that ``RASS-based surveys do not have the
sensitivity to detect clusters at $z>0.3$'' (Romer et al.\ 2000b). As
demonstrated by MACS (see the selection function shown in
Fig.~\ref{surveys} and Section~\ref{macs}), the RASS provides
unparalleled areal coverage and sufficient sensitivity to detect
hundreds of X-ray luminous clusters at $z\ga 0.3$. Whether such
systems actually exist in large numbers has, however, been the subject
of much debate. Based on very small samples, or in fact
non-detections, from serendipitous X-ray cluster surveys (the EMSS and
CfA surveys) two groups have claimed to find strong negative evolution
in the abundance of X-ray luminous clusters already at redshifts of
$z\sim 0.35$ (Henry et al.\ 1992, Vikhlinin et al.\ 1998b), in
conflict with other studies (based on the EMSS and WARPS cluster
samples) that find at best mild evolution at $z>0.5$ (Luppino \& Gioia
1995; Ebeling et al.\ 2001). As we shall show in the following, the
ROSAT All-Sky Survey holds the key to resolving this dispute which has
profound implications for our understanding of cluster evolution.

\section{The MAssive Cluster Survey (MACS)}
\label{macs}

MACS was designed to find the population of (possibly) strongly
evolving clusters, i.e., the most X-ray luminous systems at
$z>0.3$. By doing so, MACS will re-measure the rate of evolution and
test the results obtained by the EMSS and CfA cluster surveys. Unless
negative evolution is very rapid indeed, MACS will find a sizeable
number of these systems (see Section\ref{modpred}) and thus provide us
with targets for in-depth studies of the physical mechanisms driving
cluster evolution and structure formation.

In this section we give the basic X-ray selection criteria used for
MACS, derive the MACS selection function, and describe the procedure
applied to convert detect fluxes to total cluster fluxes. We then
describe the cluster identification procedure and finally present
predictions for the number of clusters expected to emerge from MACS
under the no-evolution assumption.

\subsection{X-ray selection criteria}
\label{xraycrit}

As indicated in Fig.~\ref{surveys}, MACS aims to achieve the goals
outlined above by combining the largest solid angle of any RASS
cluster survey with the lowest possible X-ray flux limit. Our survey
is based on the list of 18,811 X-ray sources contained in the RASS
Bright Source Catalogue (RASS-BSC, Voges et al.\ 1999) which has a
limiting minimal count rate of 0.05 ct s$^{-1}$ within the detect cell
and in the 0.1--2.4 keV band. Drawing from this list MACS applies the
following X-ray selection criteria:

\begin{itemize}
\item $|b|\ge 20^{\circ}$, $-40^{\circ} \le \delta {\rm (J2000)} \le
      80^{\circ}$ to ensure observability from Mauna Kea; the
      resulting geometric solid angle is 22,735 deg$^2$; 11,112
      RASS-BSC sources fall within this region
\item X-ray hardness ratio HR greater than HR$_{\rm min} = 
      \max \left[ -0.2,-0.55+\log \left( \frac{n_{\rm H}}{10^{20}\;
      {\rm cm}^{-2}}\right) \right]$ as derived from the ROSAT
      Brightest Cluster Sample (Ebeling et al.\ 1998) with the
      additional constraint that ${\rm HR}_{\rm min}<0.7$; HR is
      defined as $(h-s)/(h+s)$ where $s$ and $h$ are the PSPC
      countrates in the soft (PHA channels 11 to 41) and hard bands
      (PHA channels 52 to 201), respectively; 6,750 X-ray sources
      remain
\item $f_{\rm X} \ge 1\times 10^{-12}$ erg cm$^{-2}$
      s$^{-1}$ where $f_{\rm X}$ is the detect cell flux (see
      Section~\ref{fluxcorr}) in the 0.1--2.4 keV band; 5,654 RASS-BSC
      sources remain
\item detected net count limit of 17 photons (see Section~\ref{selfun});
      5504 sources remain
\end{itemize}

The conversion from net count rate to X-ray flux is performed using
{\sc Xspec} assuming a standard Raymond-Smith plasma spectrum, a
metallicity of 0.3, and a gas temperature $kT$ of 8 keV; we use the
Galactic $n_{\rm H}$ value from Dickey \& Lockman (1990) in the
direction of each cluster to account for absorption. The assumed X-ray
temperature of 8 keV is obtained from the $L_{\rm X}$-$kT$ relation of
White, Jones \& Forman (1997) for an X-ray luminosity of $9\times
10^{44}$ erg s$^{-1}$ (0.1--2.4 keV), typical of MACS clusters (see
Section~\ref{fluxcorr}).

We stress that we do not use the X-ray extent provided in the RASS-BSC
as a selection criterion. As shown by Ebeling and co-workers (1998,
appendix A) for the BCS ($z<0.3$) this parameter is too unreliable to
be used efficiently for the selection of cluster candidates, to the
extent that at least 25\% of all real clusters would be missed at any
given flux limit (see also Section~\ref{status}).

\subsection{X-ray selection function}
\label{selfun}

To compute the X-ray selection function, i.e., the effective solid
angle of the MACS survey as a function of X-ray flux, we need to know
the cluster detection efficiency and the depth of the RASS across our
study region. The detection algorithm used for the compilation of the
RASS-BSC is optimized for the detection of point sources and is known
to be relatively insensitive to low-surface brightness emission
(Ebeling et al.\ 1998). While this is a serious problem for the
completeness of RASS-based cluster samples at low redshift, it does
not affect MACS which, by design, targets only distant clusters. At
$z=0.3$, the limiting redshift of our survey, the canonical value of
the cluster core radius of 250 kpc corresponds to an angular size of
45 arcsec, comparable to the FWHM of the RASS point-spread function
(PSF, B\"ose 2000). Therefore the detection efficiency of distant
clusters in the RASS will not differ markedly from that of point
sources of similar X-ray flux. Hence, we can derive the effective
detection limit using all RASS-BSC sources.

\begin{figure}
\epsfxsize=0.9\textwidth 
\hspace*{5mm} \epsffile{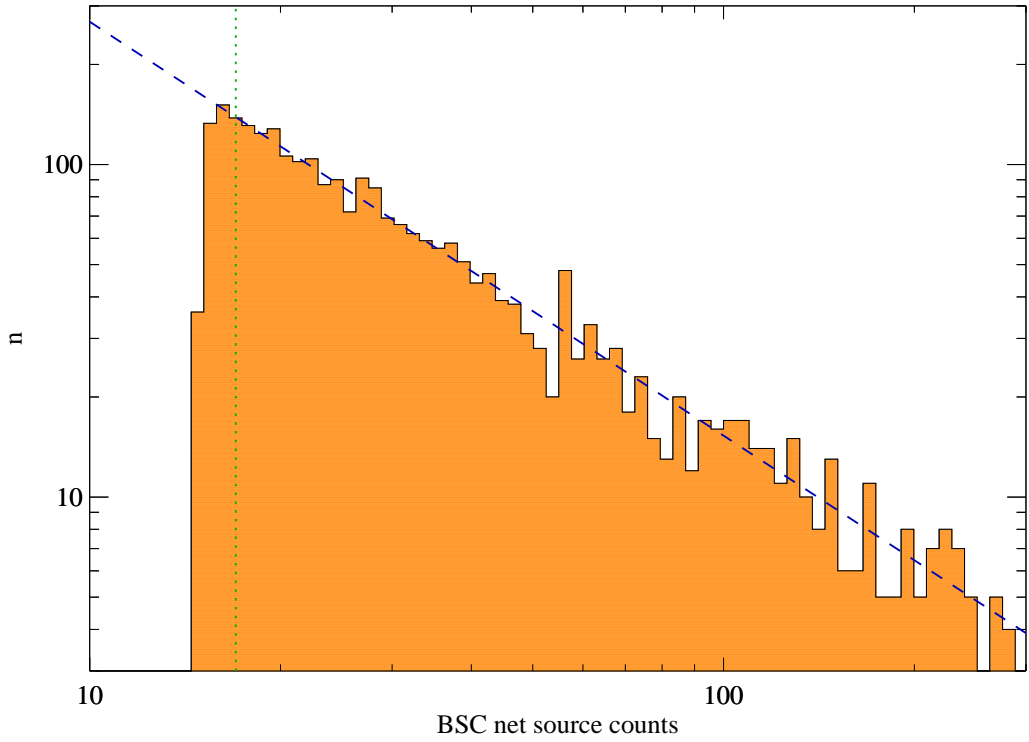}
\caption[]{\small \label{bsccts} The distribution of net detected
counts for all RASS-BSC sources with exposure times between 200 and
300 seconds. The dashed line shows the best power-law fit to the data;
the dotted line marks the completeness limit at 17 counts adopted by
us.}
\end{figure}

In Fig.~\ref{bsccts} we show the histogram of the detected net counts
of all RASS-BSC X-ray sources with exposure times between 200 and 300
seconds. The upper exposure time limit of 300s was chosen to eliminate
artificial distortions at the low-count end of the histogram due to
the presence of lower limits in the RASS-BSC in both count rate and
net counts of 0.05 ct s$^{-1}$ and 15 photons, respectively. An
additional, lower limit of 200s in exposure time was applied to create
a relatively narrow range of exposure times, thus ensuring that a
complete sample can be described by a single power law\footnote{We
stress that these exposure time cuts are applied only here to
establish the net count limit of completeness for the RASS-BSC; they
are not applied to the X-ray source list MACS is compiled from.}.
Figure~\ref{bsccts} shows that, although the RASS-BSC contains sources
with as few as 15 counts (as stated in Voges et al.\ 1999), the
catalogue is not complete to this limit. Based on a comparison with
the best-fitting power law we adopt instead a value of 17 net counts
as the completeness limit.

\begin{figure}
\epsfxsize=\textwidth 
\hspace*{0mm} \epsffile{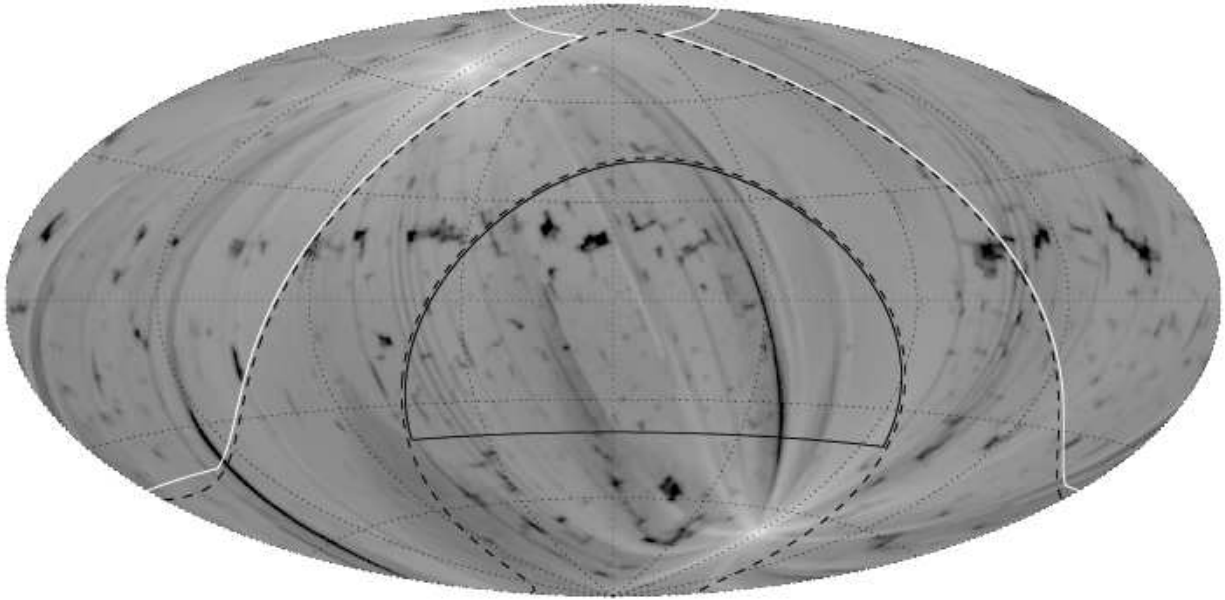}
\caption[]{\small \label{emap} RASS exposure map (Aitoff projection)
in celestial coor\-dinates 
(http://www.xray.mpe. mpg.de/rosat/survey/rass-3/sup/nx.fits.gz). The
solid white lines delineate the MACS survey area; the dashed black
lines mark the excluded 40 degree wide band centred on the Galactic
equator. The highest exposure time of several ten thousand seconds is
reached at the north ecliptic pole, the median exposure time within
the MACS survey area is 360 seconds.}
\end{figure}

\begin{figure}
\epsfxsize=0.9\textwidth 
\hspace*{5mm} \epsffile{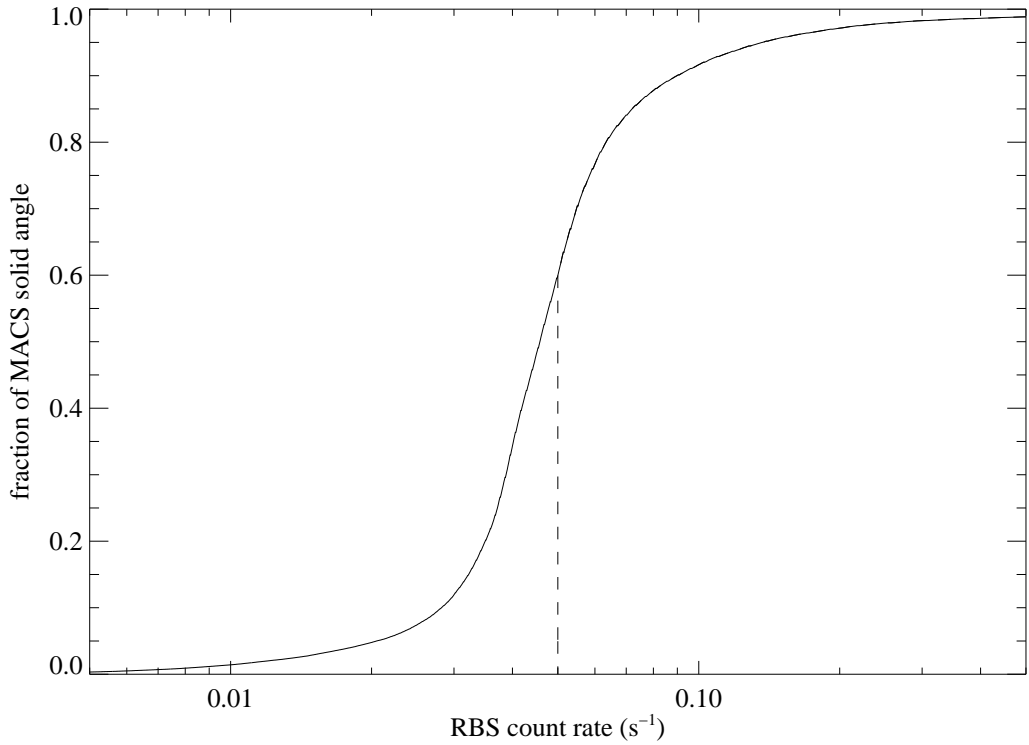}
\caption[]{\small \label{crfun} The MACS count rate selection function
corresponding to a count limit of completeness of 17 net photons in
the detection aperture. The solid line shows the fraction of the MACS
search area for which the RASS-BSC would be complete if no count rate
limit were applied.  The dashed line marks where the count rate limit
of 0.05 ct s$^{-1}$ of the RASS-BSC truncates the selection function.}
\end{figure}

\begin{figure}
\epsfxsize=0.9\textwidth 
\hspace*{5mm} \epsffile{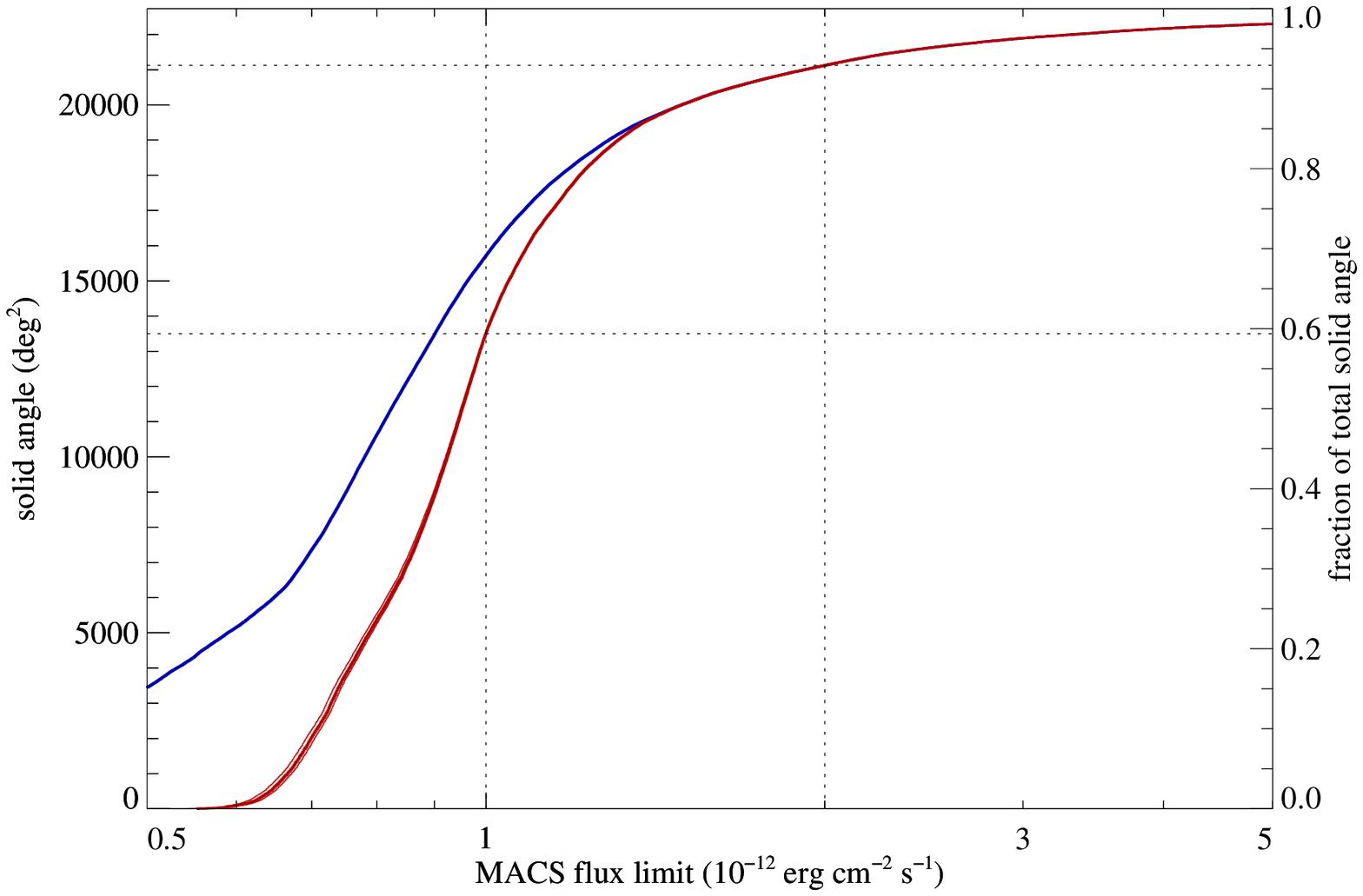}
\caption[]{\small \label{flxfun} The MACS selection function: solid
angle covered as a function of detected X-ray flux in the 0.1--2.4 keV
band. The dashed line shows the selection function attainable if
no count rate limit had been applied to the RASS-BSC.  The solid
lines show the effective MACS selection function with the count rate
limit applied and assuming cluster gas temperatures of 6, 8, and 10
keV. The dotted lines mark the detect cell flux limits, and
corresponding sky coverages, of the bright and faint MACS subsamples.}
\end{figure}

Combining the net count limit of 17 photons with the RASS exposure map
(Fig.~\ref{emap}) yields the count rate selection function, i.e., the
fractional MACS survey area for which the RASS-BSC could be complete
at a given count rate. In practice, the count rate cut at 0.05 ct
s$^{-1}$ imposed on the RASS-BSC source list truncates this function
as shown in Fig.~\ref{crfun}.  Conversion from count rate to X-ray
flux as detailed in Section~\ref{xraycrit} transforms the count rate
selection function into the desired X-ray flux selection function. As
shown in Fig.~\ref{flxfun}, and as expected from the soft energy
passband of the ROSAT PSPC, the selection function is not sensitive to
variations in the assumed X-ray temperature from 6 to 10 keV.

Based on the MACS selection function we divide the X-ray source list
compiled by applying the criteria listed in Section~\ref{xraycrit}
into an X-ray bright subset ($f_{\rm X} \ge 2\times 10^{-12}$ erg
cm$^{-2}$ s$^{-1}$) and an X-ray faint extension ($f_{\rm X} =
1-2\times 10^{-12}$ erg cm$^{-2}$ s$^{-1}$).  The bright subsample is
complete over 93\% of the geometric solid angle of our survey; when
combined with the faint extension the effective search area decreases
to 59\% of the maximal survey area of 22,735 deg$^2$. 

\subsection{Flux corrections}
\label{fluxcorr}

The X-ray fluxes derived from the RASS-BSC count rates as detailed
above are detect fluxes, i.e., they correspond to the emission
detected by the RASS-BSC detection algorithm within a specific
circular aperture. The radius of this detect cell aperture depends on
the apparent X-ray extent of the source and ranges from 5 arcmin (the
default value) to a maximal value of 16.5 arcmin.

\begin{figure}
\epsfxsize=0.9\textwidth 
\hspace*{5mm} \epsffile{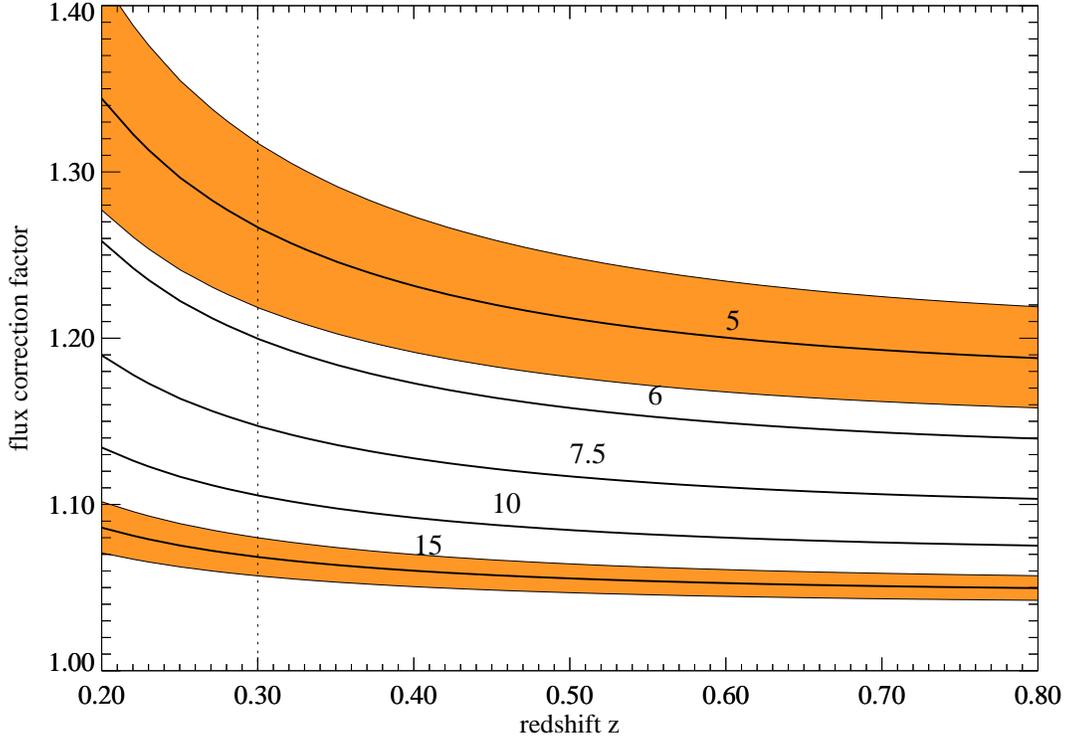}
\caption[]{\small \label{flxcorr} Flux correction factors to convert
from detect flux to total cluster flux.
The solid lines show the redshift dependence of the correction factor
for various extraction apertures with radii (in arcmin) as labeled.
For the smallest and the largest aperture the shaded regions indicate
the dependence of the flux correction factor on the assumed value of the
core radius of the emission profile (varied from 200 to 300 kpc).}
\end{figure}

To convert detect fluxes into total cluster fluxes we assume that the
intrinsic X-ray surface brightness profile follows a beta model,
$I\propto (1+r^2/r_{\rm c}^2)^{-3\beta+0.5}$ (Cavaliere \&
Fusco-Femiano 1976), with $\beta=2/3$ and core radius $r_{\rm c}=250$
kpc. We then convolve this spatial emission model with the RASS PSF
(B\"ose 2000) and compute the fraction of the observable emission that
falls within a set of circular apertures of 5, 6, 7.5, 10, and 15
arcmin radius. The resulting range of flux correction factors is shown
in Fig.~\ref{flxcorr}. Within the MACS redshift range the flux
correction factor is not a strong function of redshift; it does
depend, however, strongly on the size of the extraction aperture and,
at least for the smaller apertures, on the assumed value of $r_{\rm
c}$. While the extraction radius is known for each cluster, the core
radius is not. Next to the Poisson error of the number of directly
detected photons, the variation of the correction factor with core
radius is the second largest contributor to the uncertainty in the
total X-ray fluxes and luminosities of MACS clusters.

While the extraction radius associated with detections of nearby
clusters ($z<0.1$) is often greater than the default value (25\% of
the nearby clusters feature values greater than five arcmin), large
extraction radii become rare as the angular extent of the cluster
emission decreases with increasing cluster redshift. At $z>0.3$, the
MACS redshift range, the nominal aperture size of five arcmin radius
is used for more than 97\% of all clusters.  Using the flux correction
factor for this default extraction radius and assuming $r_{\rm c}=250$
kpc we derive limiting (minimal) total cluster luminosities of 4.7 and
$9.5\times 10^{44}$ erg s$^{-1}$ for the faint and bright MACS
subsamples at $z>0.3$; at $z>0.4$ the two subsamples contain only
clusters with luminosities in excess of 8.1 and $16.3\times 10^{44}$
erg s$^{-1}$, respectively.

\subsection{Cluster identification}
\label{ident}

The cluster identification procedure adopted for MACS involves five
steps:

\begin{enumerate}
\item Cross-correlation of the list of 5504 RASS-BSC sources with all
      objects in the SIMBAD and NED databases. Possible counterparts
      of an X-ray source are extracted within a search radius of 1
      arcmin (stars, galaxies, active galactic nuclei [AGN], QSOs) or
      3 arcmin (supernova remnant [SNRs], galaxy clusters). These
      search radii are consistent with the $3\sigma$ uncertainty of
      the RASS-BSC source positions of 1 arcmin (98\% limit of error
      distribution) for point sources and 2 arcmin (98\% limit) for
      extended sources, and account for an additional uncertainty of
      about 1 arcmin in the positions of catalogued supernova remnants
      and clusters of galaxies.
\item Visual inspection of Digitized Sky Survey (DSS) images
      (second generation where available). The size of these
      images is $5\times 5$ arcmin$^2$ corresponding to at least $1.65
      \times 1.65$ Mpc$^2$ within the redshift range of our survey
      ($z>0.3$).
\item Search for extremely blue ($O-E<1.3$) or red ($O-E>2$)
      counterparts in the APM (Automated Plate Measuring machine;
      Irwin, Maddox \& McMahon 1994) object catalogue to tentatively
      identify stars, AGN, and BLLac objects. Only objects within 25''
      of the RASS-BSC X-ray position are considered. The quoted colour
      and angular separation thresholds correspond to 95\% confidence
      limits for identifications with these types of objects obtained
      from cross-correlations of the APM catalogue with known AGN and
      stars.
\item CCD imaging in the R (bright source list) or I band (faint
      source list) of all X-ray sources without (or with ambiguous)
      identifications as well as of all possibly distant ($z\ga 0.2$)
      cluster candidates with the University of Hawaii's 2.2m
      telescope.  At exposure times of $3 \times 2$ min in R and
      $3\times 3$ min in I these imaging observations are deep enough
      to unambiguously detect rich clusters out to $z\sim
      0.8$\footnote{ The WARPS team discovered the rich clusters
      ClJ0152.7$-$1357 ($z=0.833$, Ebeling et al.\ 2000b) and
      ClJ1226.9+3332 ($z=0.888$, Ebeling et al.\ 2001) in the first
      of three 4 min I band exposures taken with the same
      instrumentation at the UH2.2m telescope as is used by us for
      MACS. The mentioned two WARPS clusters constitute a complete
      sample at this redshift.}
\item Spectroscopic observations with the UH2.2m and Keck 10m
      telescopes of all confirmed clusters with estimated redshifts of
      $z\ga 0.2$.
\end{enumerate}

For an RASS-BSC source to be flagged as a non-cluster before CCD
images are obtained, the cross-correlation with Galactic and
extragalactic object catalogues has to yield an unambiguous
non-cluster identification that is supported by the appearance of the
field in the DSS finders, as well as by the APM colour (where
available) of the counterpart.  Typical examples of such obvious
identifications are bright stars and nearby galaxies (with and without
nuclear activity).  We stress that DSS finders are obtained and
examined for all RASS-BSC sources meeting the initial X-ray selection
criteria (Section~\ref{xraycrit}), and that we do proceed to CCD
imaging in spite of the presence of a listed non-cluster counterpart
if, for instance, a catalogued QSO is not clearly visible in the DSS
image or if, in addition to the QSO, an overdensity of faint objects
is apparent in our finders. 

Unless a RASS-BSC source has been firmly identified as a non-cluster,
or as a cluster at $z<0.2$ (where $z$ can be a measured or estimated
redshift), CCD images of the source over a $7.5\times 7.5$ arcmin$^2$
field-of-view are obtained with the University of Hawaii's 2.2m
telescope. Since MACS clusters, in contrast to the majority of the
systems detected in serendipitous cluster surveys, are by design and
without exception very X-ray luminous (Section~\ref{fluxcorr}) they
are usually also optically rich and thus obvious even in shallow CCD
images. All distant clusters ($z_{\rm est}>0.2$) confirmed by imaging
observations are subsequently targeted in spectroscopic observations
where we obtain redshifts of at least two cluster members, one of them
the apparent brightest cluster member.

\begin{figure}
\epsfxsize=\textwidth 
\hspace*{0mm} \epsffile{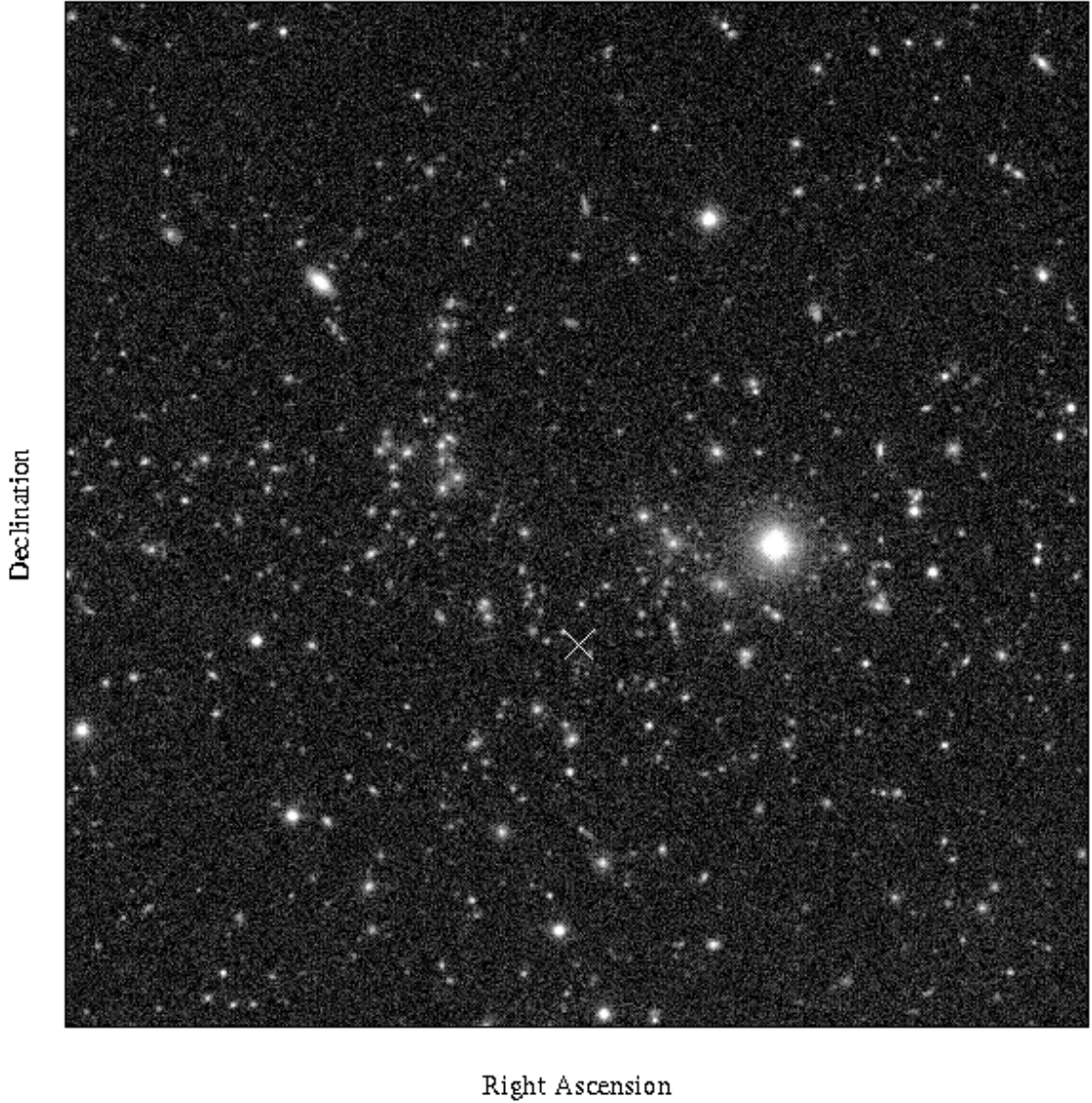}
\caption[]{\small \label{macsvri} Colour image ($5\times 5$
arcmin$^2$) of a newly discovered MACS cluster at $z=0.453$, based on
$3\times 4$min exposures in each of the V, R and I bands with the
University of Hawaii's 2.2m telescope. The RASS-BSC X-ray position is
30'' south of the image center.  We obtain images like this one for
all MACS clusters with spectroscopic redshifts of $z>0.3$ to allow the
optical richness of these systems to be assessed, to efficiently
select cluster galaxies for multi-object spectroscopy, and to identify
unusually red or blue objects that might be X-ray contaminants. }
\end{figure}

A systematic effect that is difficult to quantify is the impact of
X-ray contamination on our sample. We cannot rule out that we may have
included a small number of clusters at $z>0.3$ that are significantly
contaminated by X-ray point sources and would fall below our flux
limit if the non-diffuse emission were subtracted. We attempt to
identify possibly contaminated clusters by obtaining deeper ($3\times
4$ min) optical images in each of three passbands (V, R, I) of all
MACS clusters with spectroscopic redshifts of $z>0.3$.
Fig.~\ref{macsvri} shows such a colour image (of a newly discovered
MACS cluster at $z=0.453$) and illustrates how the optical richness of
MACS clusters allows an unambiguous identification already from
relatively shallow CCD images. A (by MACS standards) low optical
richness of a system in these colour images is one possible indicator
of contamination, as is the presence of unusually red or blue objects
close to the X-ray position. In future follow-up work we shall attempt
to obtain spectra of potential contaminants identified in this
manner. However, ultimately we will be not be able to quantify the
level of X-ray contamination until deeper pointed X-ray observations
of all MACS clusters have been performed.

While we believe to have taken all feasible precautions against
missing distant clusters, the above procedure (or any other) can never
be failsafe. Albeit unlikely, a distant cluster might be obscured by a
bright star which we accepted as the X-ray counterpart.
Alternatively, a very distant cluster which is not visible on the DSS
finder can be missed if an acceptable optical counterpart to the X-ray
source is present in the foreground (catalogued AGN or QSO at lower
redshift). While, in both of these examples, the eventually accepted
identification is likely to contribute to the observed X-ray emission,
we can not rigorously rule out that we have missed a small number of
distant clusters above our X-ray flux limit. As for all cluster
surveys, the size of the cluster sample emerging from MACS, as well as
all volume-normalized quantities derived from it, should thus be
considered to represent lower limits. 

\subsection{No-evolution prediction}
\label{modpred}

A prediction for the size of the final MACS sample under the
no-evolution assumption can be obtained by folding the local cluster
X-ray luminosity function (XLF) as measured from the ROSAT Brightest
Cluster Sample (Ebeling et al.\ 1997) through the MACS selection
function shown in Fig.~\ref{flxfun}. In this process, we use our usual
assumptions to convert from X-ray luminosity to flux and from total
cluster flux to detect cell flux (see Sections~\ref{xraycrit} and
\ref{fluxcorr}) and integrate the cluster XLF out to $z=1$.

\begin{figure}
\epsfxsize=0.9\textwidth 
\hspace*{5mm} \epsffile{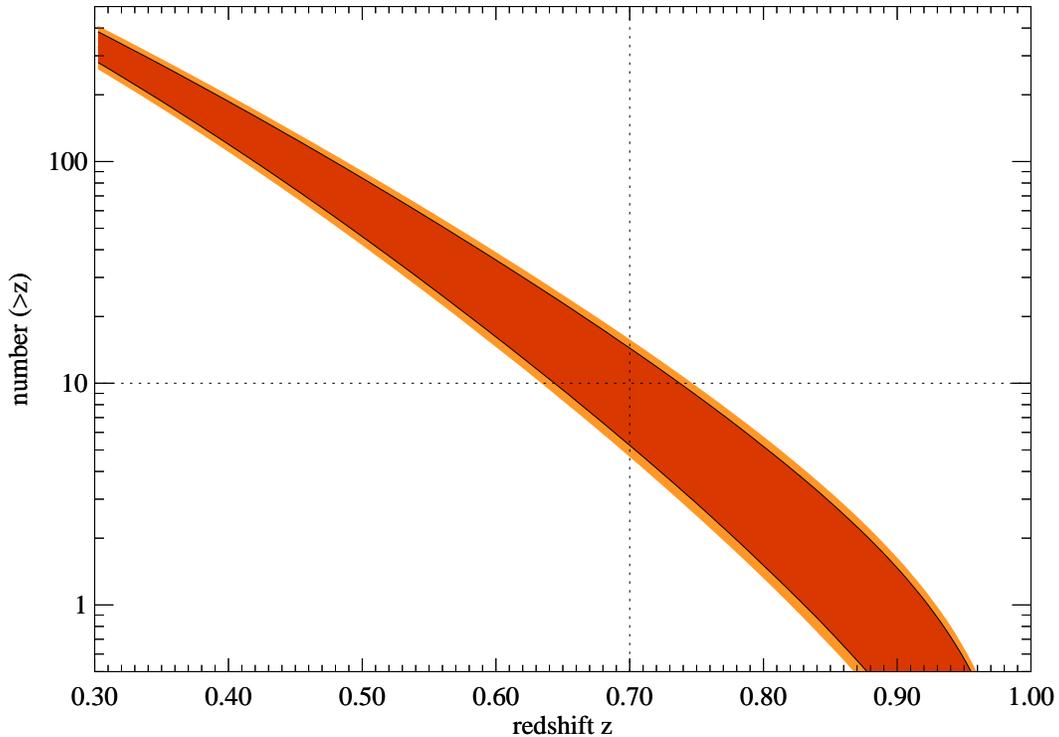}
\caption[]{\small \label{pred} Number of clusters above redshift $z$
predicted to emerge from MACS under the no-evolution assumption.  We
use the local cluster X-ray luminosity function (XLF) from the BCS
(Ebeling et al.\ 1997). The uncertainty in the prediction indroduced
by the errors in the Schechter function parameterization of the BCS
XLF is represented by the dark shading; the light shading shows the
additional effect of varying the core radius between 200 and 300 kpc.
As shown by the dotted lines, of the order of ten clusters at $z>0.7$
are expected if the XLF does not evolve out to $z=1$.}
\end{figure}

The resulting model prediction is shown in Fig.~\ref{pred}.  Although
the uncertainties introduced by the errors in the Schechter function
parameterization of the local XLF ($z<0.3$) are considerable, it is
safe to say that about 300 clusters are expected to emerge from MACS
if there is no evolution in the cluster XLF out to $z=1$. Only at
$z>0.7$ would the number of MACS clusters approach or fall below about
ten, thus entering the Poisson regime.

These numbers are sufficiently high for us to be confident that MACS
will produce not only the largest sample of massive, distant clusters
compiled to date (a relative statement), but also a sizeable one in
absolute terms, even in the presence of strong negative evolution.

\section{MACS: Status as of December 2000}
\label{status}

\begin{figure}
\epsfxsize=\textwidth
\hspace*{0mm} \epsffile{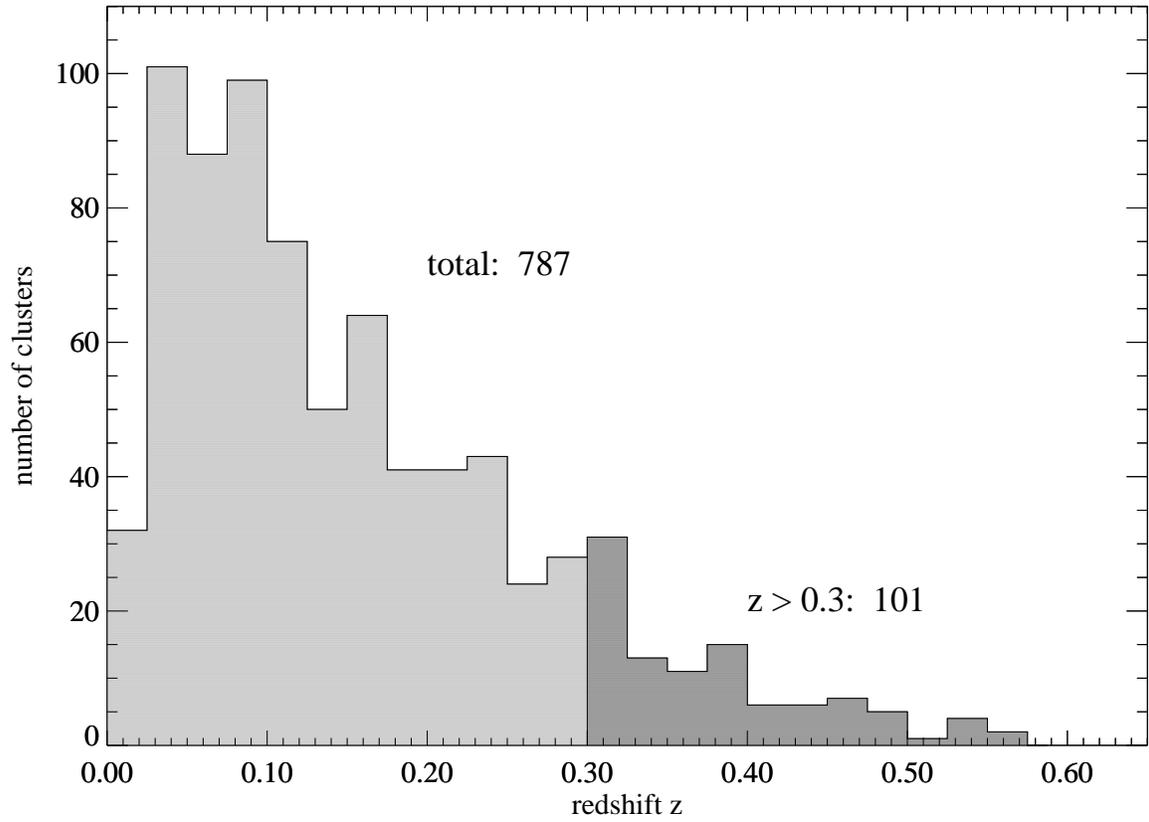}
\caption[]{\small \label{zdist} The redshift distribution of the 787
clusters identified in the MACS project to date. The 101 clusters at
$z>0.3$ that form the preliminary MACS sample are highlighted. All
clusters have spectroscopic redshifts.}
\end{figure}

As part of the procedure described in detail in Section~\ref{ident} we
have, so far, obtained 349 CCD images of MACS cluster candidates and
measured redshifts for 131 clusters confirmed by the imaging
observations. As of December 2000, we have identified more than 850
clusters of galaxies at all redshifts; Fig.~\ref{zdist} shows the
redshift distribution of the 787 systems with spectroscopic
redshifts. As a by-product, MACS has thus already delivered by far the
largest X-ray selected cluster catalogue to emerge from the RASS to
date.

The redshift distribution shown in Fig.~\ref{zdist} is skewed toward
high redshifts because our spectroscopic follow-up observations target
exclusively systems with $z_{\rm{est}}>0.2$. Up to December 2000 101
clusters were found to have $z>0.3$; a further 37 clusters confirmed
in imaging observations and with $z_{\rm est}>0.2$ still await
spectroscopic confirmation. Of the 101 clusters in the preliminary
MACS sample only 29 were previously known. These 29 hail from a wide
variety of projects, including the optically selected GHO sample
(Gunn, Hoessel \& Oke 1986), the Abell catalogue (Abell, Corwin \&
Olowin 1989) and the X-ray selected EMSS\footnote{The less than a
handful EMSS clusters rediscovered by MACS constitute the largest
statistically complete previous cluster sample in this redshift and
X-ray luminosity range.} (Gioia \& Luppino 1994) and BCS (Ebeling et
al.\ 1998, 2000) cluster samples.

\begin{figure}
\epsfxsize=\textwidth
\hspace*{0mm} \epsffile{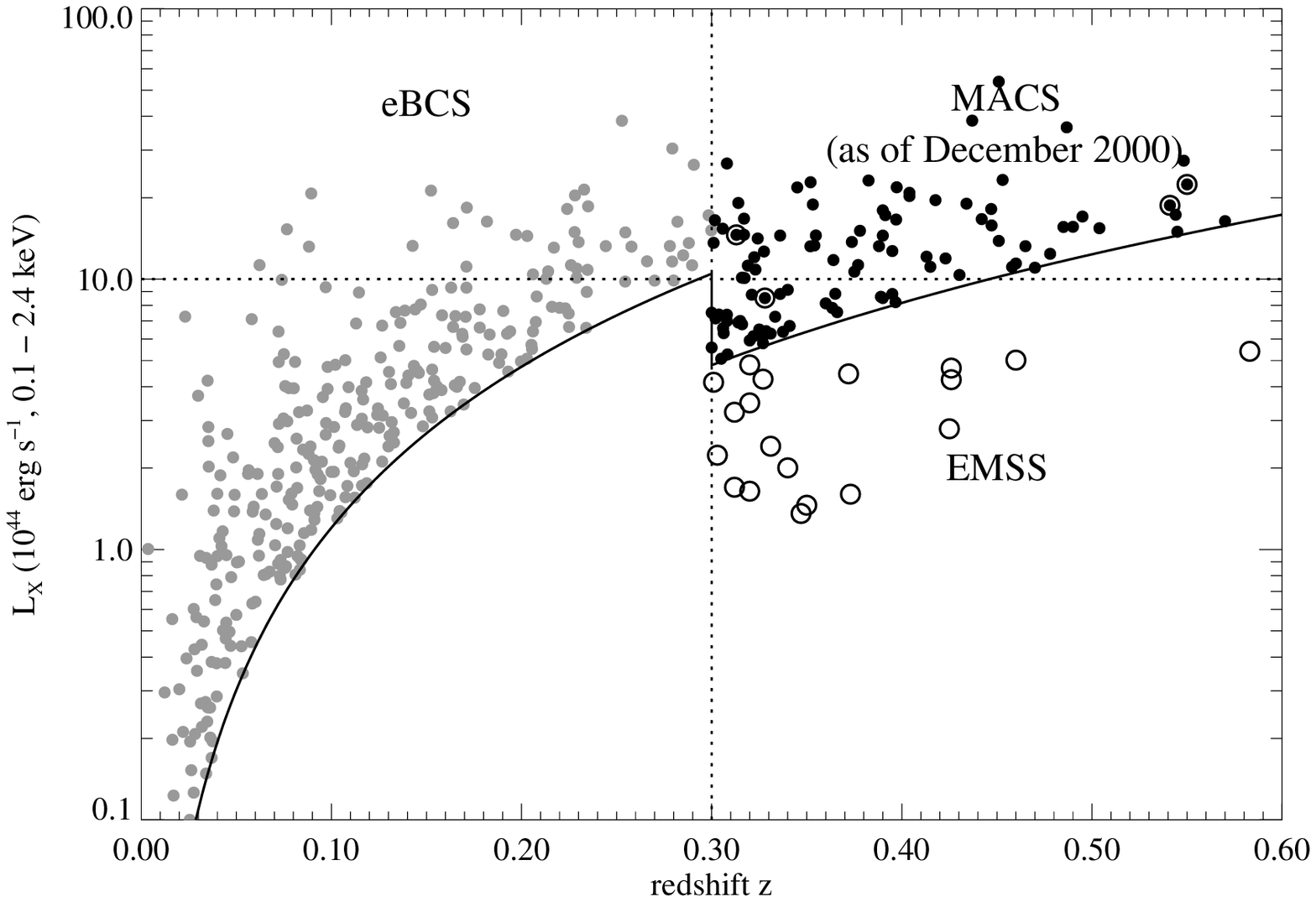}
\caption[]{\small \label{lxz} The luminosity--redshift distribution of
the extended BCS (Ebeling et al.\ 1998, 2000) at $z<0.3$ and of the
preliminary MACS sample (101 clusters) at $z>0.3$. Also plotted (open
circles) are the loci of the 23 EMSS clusters at $0.3<z<0.6$ (Henry et
al.\ 1992).  The solid line marks the flux limits of the BCS and MACS
surveys; the dashed line shows the flux limit of the X-ray bright MACS
subsample.  By design MACS finds the high-redshift counterparts of the
most X-ray luminous (and best studied) clusters in the local
universe.}
\end{figure}

Figure~\ref{lxz} shows the X-ray luminosity--redshift distribution of
our preliminary sample at $z>0.3$, compared to the one for the BCS at
$z<0.3$, and the one for the EMSS at $0.3<z<0.6$. Note how MACS
extends the redshift baseline for studies of the most X-ray luminous
clusters ($L_{\rm X}\sim 1\times 10^{45}$ erg s$^{-1}$) from $z\la
0.3$ to $z\la 0.6$, and how MACS clusters are, in general, much more
X-ray luminous than EMSS clusters. In fact only six EMSS clusters come
close to the X-ray luminosities sampled by MACS at $z>0.3$; four of
these are rediscovered by us (MS2137.3$-$2353, MS1358.4+6245,
MS0451.6$-$0305, and MS0015.9+1609), the other two (MS0353.6$-$3642
and MS1008.1$-$1224) lie just below our flux limit. MACS thus contains
already more than 15 times more clusters in this cosmologically most
important region of the $L_{\rm X}$-$z$ plane than the EMSS, providing
us for the first time with a sizeable and statistically robust sample
for studies of the properties of the high-redshift counterparts of the
most massive local clusters.

\begin{figure}
\epsfxsize=\textwidth
\hspace*{0mm} \epsffile{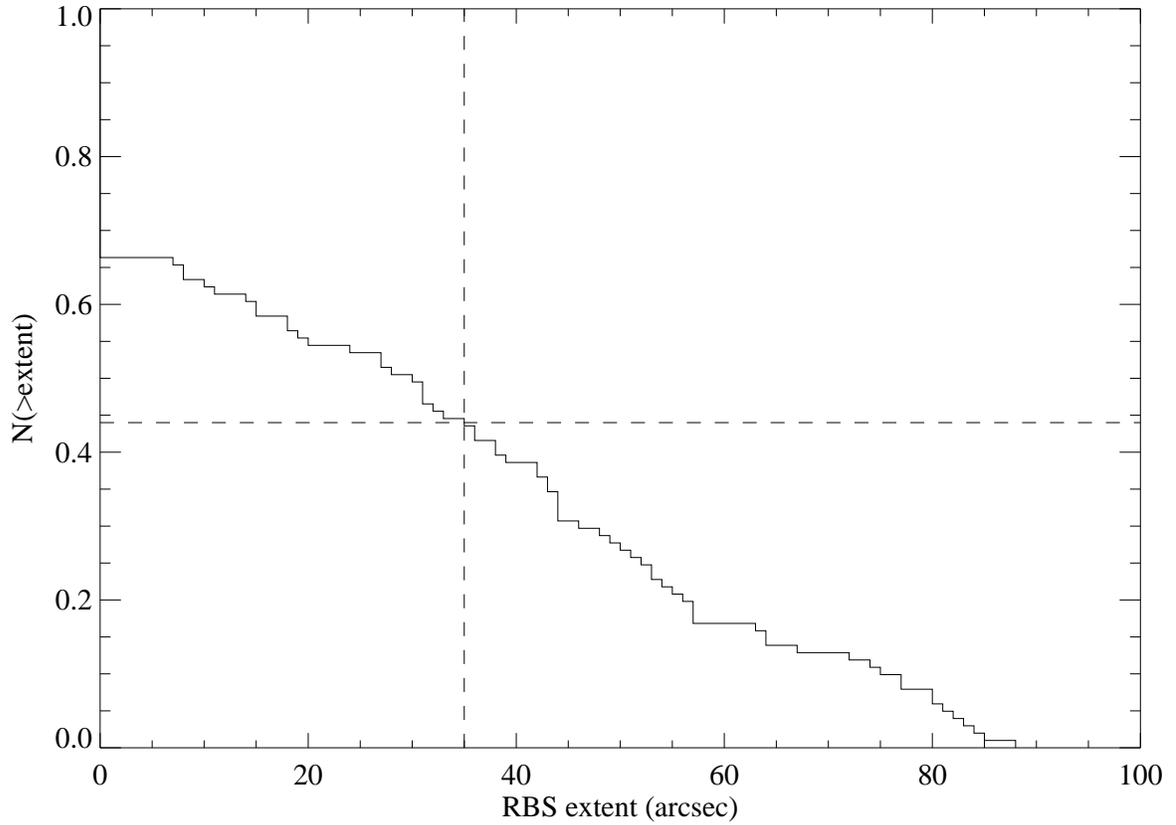}
\caption[]{\small \label{macsext} The cumulative RASS-BSC extent
distribution of the 101 MACS clusters in our preliminary sample. The
dashed lines mark the extent threshold of 35 arcsec above which a
source can be considered to be genuinely extended according to Ebeling
et al.\ (1993), and the completeness (44\%) of the MACS sample that
would have resulted if this extent threshold had been used as an X-ray
selection criterion.}
\end{figure}

We emphasize again that the completeness of this sample, compiled from
RASS data, hinges critically upon our ignoring the RASS-BSC extent
parameter.  As shown in Fig.~\ref{macsext} 34\% of all MACS clusters
at $z>0.3$ are classified as X-ray point sources by the RASS-BSC
detection algorithm.  Based on a comparison of the extent values
assigned by the RASS-BSC algorithm to detections of Abell clusters and
to a control set of random RASS sources, Ebeling and coworkers (1993)
find that extent values below 35 arcsec are in general spurious. If
this threshold value had been adopted for MACS, our survey would have
missed more than half of the 101 clusters in our preliminary sample.

\section{Summary}

We describe the design and status of the MAssive Cluster Survey
(MACS), the first X-ray cluster survey aimed at the compilation of a
large, statistically complete sample of exclusively X-ray luminous
($L_{\rm X}\ga 5\times 10^{44}$ erg s$^{-1}$, 0.1--2.4 keV), distant
($z>0.3$) clusters. The systems targeted by our survey are the rarest,
most massive clusters whose evolution places the tightest constraints
on the physical and cosmological parameters of structure formation on
cluster scales.

Based on the ROSAT Bright Source Catalogue of RASS detections, MACS
uses the spectral hardness of the X-ray emission and the X-ray flux in
the detect aperture to select 5504 X-ray sources in a search area of
22,735 deg$^2$. A comprehensive identification programme has so far
led to the discovery of more than 800 clusters at all redshifts;
imaging and spectroscopic follow-up observations have confirmed 101
clusters at $z>0.3$. MACS has thus already more than tripled the
number of massive, distant clusters known; compared to the EMSS sample
our current preliminary sample represents an improvement in size of a
factor of 15 in the MACS redshift and luminosity range.

Under the no-evolution assumption, MACS is expected to uncover up to,
and perhaps more than, 300 clusters at $z>0.3$. However, if evolution
is strong and negative, the total sample could comprise as few as as
100 clusters.  In any case MACS will increase greatly the number of
distant, massive clusters known and, hopefully, lead to similarly
impressive improvements in our understanding of the properties and
evolution of these most massive collapsed entities in the universe.

\acknowledgements

We thank the telescope time allocation committee of the University of
Hawai`i for their generous support of the MACS optical follow-up
program. HE gratefully acknowledges financial support from NASA LTSA
grant NAG 5-8253. ACE thanks the Royal Society for financial support.
This research has made use of the NASA/IPAC Extragalactic Database
(NED) which is operated by the Jet Propulsion Laboratory, California
Institute of Technology, under contract with the National Aeronautics
and Space Administration. The Digitized Sky Surveys were produced at
the Space Telescope Science Institute under US Government grant NAG
W-2166. The images of these surveys are based on photographic data
obtained using the Oschin Schmidt Telescope on Palomar Mountain and
the UK Schmidt Telescope.

\end{document}